\documentclass[twocolumn,floatfix,showpacs,showkeys,preprintnumbers,nofootinbib,superscriptaddress]{revtex4}
\usepackage[utf8]{inputenc}
\usepackage[sort&compress]{natbib}
\usepackage{ulem}
\usepackage{bm}
\usepackage{times}
\usepackage{amssymb,amsbsy,amsmath,amsfonts}
\usepackage{graphicx}
\usepackage{float}
\usepackage{color}
\usepackage{morefloats}
\usepackage{rotating}
\usepackage{srcltx}
\usepackage{slashed}
\usepackage{subfigure}
\usepackage{multirow}
\usepackage{verbatim}
\usepackage{hyperref}
\usepackage{tabularx}

\begin{document}

\title{Electromagnetic form factors of $\Sigma^{+}$ and $\Sigma^{-}$ in the vector meson dominance model }

\author{Zhong-Yi Li}
\affiliation{Institute of Modern Physics, Chinese Academy of Sciences, Lanzhou 730000, China}
\affiliation{School of Nuclear Science and Technology, University of Chinese Academy of Sciences, Beijing 101408, China}
\author{Ju-Jun Xie}\email{xiejujun@impcas.ac.cn}
\affiliation{Institute of Modern Physics, Chinese Academy of Sciences, Lanzhou 730000, China}
\affiliation{School of Nuclear Science and Technology, University of Chinese Academy of Sciences, Beijing 101408, China}
\affiliation{School of Physics and Microelectronics, Zhengzhou University, Zhengzhou, Henan 450001, China }

\date{\today}
\begin{abstract}

Based on the recent measurements of the $e^{+} e^{-} \to
\Sigma^{+} \bar{\Sigma}^{-}$ and $e^{+} e^{-} \to \Sigma^{-}
\bar{\Sigma}^{+}$ processes by BESIII collaboration, the
electromagnetic form factors of the hyperon $\Sigma^+$ and
$\Sigma^-$ in the timelike region are investigated by using the
vector meson dominance model, where the contributions from the
$\rho$, $\omega$, and $\phi$ mesons are taken into account. The
model parameters are determined from the BESIII experimental data on
the timelike effective form factors $|G_{\rm eff}|$ of $\Sigma^{+}$
and ${\Sigma}^{-}$ baryons for center-of-mass energy from 2.3864 to
3.02 GeV. It is found that we can provide quantitative descriptions
of available data as few as one adjustable model parameter. We then progress to an analysis of the electromagnetic form factors in the spacelike region and
evaluate the spacelike form factors of the hyperons
$\Sigma^+$ and $\Sigma^-$. The obtained electromagnetic form factors
of the $\Sigma^+$ and $\Sigma^-$ baryons are comparable with other
model calculations.

\end{abstract}
\maketitle

\section{Introduction}

The study of the structure of hyperons which contain strange quark
through their electromagnetic form factors (EMFFs) is crucial for
deep understanding the nonperturbative quantum chromodynamics (QCD)
effects that quarks are bounded in
baryons~\cite{Yang:2019mzq,Ramalho:2019koj,Yang:2020rpi,Ablikim:2020kqp,Haidenbauer:2020wyp}.
As discussed in
Refs.~\cite{Geng:2008mf,Green:2014xba,Brodsky:1974vy}, the EMFFs are
crucial experimental observables of baryons which are intimately
related to their internal structure and dynamics. In the last few
decades, there are great progress in the study of baryon EMFFs,
especially for the nucleon both in
timelike~\cite{Akhmetshin:2015ifg,Andreotti:2003bt,Antonelli:1998fv,Ablikim:2019eau,Bardin:1994am,Bisello:1983at,Ambrogiani:1999bh,Aubert:2005cb,Lees:2013uta,Lees:2013ebn,Ablikim:2015vga,Armstrong:1992wq}
and spacelike
regions~\cite{Zhu:2001md,Passchier:2001uc,Gayou:2001qd,Madey:2003av,Warren:2003ma,Becker:1999tw,Bermuth:2003qh,Golak:2000nt,Gayou:2001qt,Ostrick:1999xa,Rohe:1999sh}.
For example, in Ref.~\cite{Chen:2007sa} the nucleon form factors and
the $N$-$\Delta(1232)$ transitions were theoretically investigated
in a framework of constituent quark model, where the effect of $\pi$
meson clound is also considered. While for the case of hyperons,
since they cannot be targets, the hyperon EMFFs in the spacelike
region are hardly to be measured by experiments. In the timelike
region, the BESIII collaboration investigated the $\Sigma$ hyperon
EMFFs from the reaction $e^{+} e^{-} \to \Sigma^{+}
\bar{\Sigma}^{-}$ and  $e^{+} e^{-} \to \Sigma^{-}
\bar{\Sigma}^{+}$~\cite{Ablikim:2017wlt,Ablikim:2020kqp}. From the
determined high precision Born cross sections of these two above
reactions for center-of-mass energy from 2.3864 to 3.02 GeV, the
effective form factors $|G_{\rm eff}|$ of $\Sigma^{+}$ and
${\Sigma}^{-}$ and also the ratios of $\Sigma^+$ electric and
magnetic form factors $|G_{E}/G_{M}|$, are
obtained~\cite{Ablikim:2020kqp}.

Very recently, the EMFFs of hyperons are investigated in the
timelike region through $e^+e^- \to Y\bar{Y}$ ($Y$ is the hyperon)
reaction~\cite{Haidenbauer:2020wyp}, where the $Y\bar{Y}$ final
state interactions are taken into account, and the EMFFs in the
timelike region are calculated for the $\Lambda$, $\Sigma$, and
$\Xi$ hyperons based on one-photon approximation for the elementary
reaction mechanism. In Ref.~\cite{Ramalho:2019koj}, the timelike
EMFFs of hyperons are well reproduced within a relativistic quark
model. In addition, the $\Sigma$ hyperon EMFFs have been calculated
within Lattice QCD~\cite{Lin:2008mr}, light cone sum rule
(LCSR)~\cite{Liu:2009mb}, and chiral perturbative theory
(ChPT)~\cite{Kubis:2000aa}. In addition to these, the vector meson dominance
(VMD) model is a very successful approach to study the nucleon
electromagnetic form factors both in spacelike and timelike
regions~\cite{Iachello:1972nu,Iachello:2004aq,Bijker:2004yu}. Within
a modified VMD model, in Ref.~\cite{Yang:2019mzq} the EMFFs of
$\Lambda$ hyperon were studied in the timelike region from $e^+e^-
\to \Lambda \bar{\Lambda}$ reaction, where the contributions from
$\phi$ and $\omega$ mesons, and their excited states are included.
It was found that the VMD model can simultaneously describe the
effective form factor and also the electromagnetic form factor ratio
of the $\Lambda$ hyperon.

In this work, based on the recently measurements of the $e^{+} e^{-}
\to \Sigma^{+} \bar{\Sigma}^{-}$ and $e^{+} e^{-} \to \Sigma^{-}
\bar{\Sigma}^{+}$ reactions, we aim to determine the parameters of
VMD model by fitting them to the experimental data of $|G_{\rm
eff}|$ of $\Sigma^{+}$ and ${\Sigma}^{-}$. We have included the
contributions from the $\rho$, $\omega$ and $\phi$ mesons. Then the
ratios $|G_{E}/G_{M}|$ of $\Sigma^+$ are estimated with the model
parameters, which are comparable with other theoretical
calculations.

This article is organized as follows: formalism of $\Sigma$ hyperon
form factors in VMD model are shown in the following section.  In
Sec. III, we introduce effective form factors and the method that
analytically continue the expressions of the form factors from the
timelike region to the spacelike region. Numerical results of the
timelike form factors of $\Sigma^{+}$ and $\Sigma^{-}$ hyperon and
the ratios $|G_{E}/G_{M}|$ of $\Sigma^{+}$ and $\Sigma^{-}$ are
presented. In Sec. IV, followed by a short summary in the last
section.

\section{Theoretical formalism}

We will study the EMFFs of $\Sigma^+$ and $\Sigma^-$ within the VMD
model. As in Ref.\cite{Yang:2019mzq}, we first introduce the
electromagnetic current of $\Sigma$ hyperon with spin-$1/2$ in terms
of the Dirac form factor $ F_{1}(Q^{2})$ and Pauli form factors
$F_{2}(Q^{2})$ as

\begin{equation}
J^{\mu}=\gamma^{u}F_{1}(Q^{2})+ {\rm i}\frac{\sigma^{\mu \nu}q_{\nu}}{2m_{\Sigma}}F_ {2}(Q^2),
\end{equation}
where $F_1$ and $F_2$ are functions of the squared momentum transfer
$Q^2 = -q^2$. In the VMD model, the Dirac and Pauli form factors are
parametrized into two parts. One is the intrinsic three-quark
structure described by the form factor
$g(Q^{2})$~\cite{Bijker:2004yu}, the other one is the meson cloud,
which is used to describe the interaction between the bare baryon
and the photon through the intermediate isovector $\rho$ meson and
isoscalar $\omega$ and $\phi$ mesons~\cite{Iachello:1972nu}.
Following
Refs.~\cite{Iachello:1972nu,Iachello:2004aq,Bijker:2004yu}, $F_1$
and $F_2$ can be decomposed as $F_{i}=F_{i}^{S}+F_{i}^{V}$, with
$F_{i}^{S}$ and $F_{i}^{V}$ the isoscalar and isovector components
of the form factors, respectively. The Dirac and Pauli form
factors of $\Sigma^{+}$ and  $\Sigma^{-}$ are easily obtained
without considering the total decay widths of the vector mesons, as
follows
\begin{widetext}
\begin{eqnarray}
F_{1 \Sigma^{+}}^{S}\left(Q^{2}\right)&=&\frac{1}{2} g_{1}\left(Q^{2}\right)\left[\left(1-\beta_{\omega}-\beta_{\phi}\right)+\beta_{\omega} \frac{m_{\omega}^{2}}{m_{\omega}^{2}+Q^{2}}\right.
\left.+\beta_{\phi} \frac{m_{\phi}^{2}}{m_{\phi}^{2}+Q^{2}}\right],  \label{eq:f1sigmaps} \\
F_{1 \Sigma^{+}}^{V}\left(Q^{2}\right)&=&\frac{1}{2} g_{1}\left(Q^{2}\right)\left[\left(1-\beta_{\rho}\right)+\beta_{\rho} \frac{m_{\rho}^{2}}{m_{\rho}^{2}+Q^{2}}\right],  \label{eq:f1sigmapv}\\
F_{2 \Sigma^{+}}^{S}\left(Q^{2}\right)&=&\frac{1}{2} g_{1}\left(Q^{2}\right)\left[\left(2\mu_{\Sigma^+} - 2 -\alpha_{\phi}-\alpha_{\rho}\right) \frac{m_{\omega}^{2}}{m_{\omega}^{2}+Q^{2}}+\alpha_{\phi} \frac{m_{\phi}^{2}}{m_{\phi}^{2}+Q^{2}}\right],  \label{eq:f2sigmaps}\\
F_{2 \Sigma^{+}}^{V}\left(Q^{2}\right)&=&\frac{1}{2}
g_{1}\left(Q^{2}\right)\left[\alpha_{\rho}
\frac{m_{\rho}^{2}}{m_{\rho}^{2}+Q^{2}}\right],
\label{eq:f2sigmapv}\\
F_{1 \Sigma^{-}}^{S}\left(Q^{2}\right)&=&\frac{1}{2} g_{2}\left(Q^{2}\right)\left[\left(-1-\beta_{\omega}-\beta_{\phi}\right)+\beta_{\omega} \frac{m_{\omega}^{2}}{m_{\omega}^{2}+Q^{2}}\right.
\left.+\beta_{\phi} \frac{m_{\phi}^{2}}{m_{\phi}^{2}+Q^{2}}\right],  \label{eq:f1sigmamfs}\\
F_{1 \Sigma^{-}}^{V}\left(Q^{2}\right)&=&\frac{1}{2} g_{2}\left(Q^{2}\right)\left[\left(-1-\beta_{\rho}\right)+\beta_{\rho} \frac{m_{\rho}^{2}}{m_{\rho}^{2}+Q^{2}}\right],  \label{eq:f1sigmamfv}\\
F_{2 \Sigma^{-}}^{S}\left(Q^{2}\right)&=&\frac{1}{2} g_{2}\left(Q^{2}\right)\left[\left(2\mu_{\Sigma^-} + 2 -\alpha_{\phi}-\alpha_{\rho}\right) \frac{m_{\omega}^{2}}{m_{\omega}^{2}+Q^{2}}+\alpha_{\phi} \frac{m_{\phi}^{2}}{m_{\phi}^{2}+Q^{2}}\right],  \label{eq:f2sigmamfs}\\
F_{2 \Sigma^{-}}^{V}\left(Q^{2}\right)&=&\frac{1}{2}
g_{2}\left(Q^{2}\right)\left[\alpha_{\rho}
\frac{m_{\rho}^{2}}{m_{\rho}^{2}+Q^{2}}\right],
\label{eq:f2sigmamfv}
\end{eqnarray}
\end{widetext}
where we take the intrinsic form factor $g(Q^2)$ as a dipole form
\begin{eqnarray}
g_{1}(Q^{2}) &=& (1+\gamma_{1}Q^{2})^{-2} , \\
g_{2}(Q^{2}) &=& (1+\gamma_{2}Q^{2})^{-2} .
\end{eqnarray}
This form is consistent with $g=1$ at $Q^2 = 0$, and the free model
parameters $\gamma_1$ and $\gamma_2$ will be determined by fitting
them to the experimental data.

On the other hand, the observed electric and magnetic form factors
$G_{E}$ and $G_{M}$ can be expressed in terms of Dirac and Pauli
form factors $F_1$ and $F_2$ by,
\begin{eqnarray}
G_E (Q^2)\! &=&\! F_1 - \tau F_2 = F^S_1 + F^V_1 - \tau (F^S_2 + F^V_2),\\
G_M (Q^2)\! &=&\! F_1 + F_2 = F^S_1 + F^V_1 + F^S_2 + F^V_2,
\end{eqnarray}
where $\tau = \frac{Q^2}{4M^2_\Sigma}$ in this work. At $Q^2$ = 0,
$G_M(Q^2)$ defines the value of the magnetic moment of the $\Sigma$
hyperon, $\mu_\Sigma = G_M(0)$, in natural unit, i.e.,
$\hat{\mu}_\Sigma = e/(2M_\Sigma)$. For easily compare magnetic
moments of different particles masses it is usual to express
magnetic moments in terms of $\hat{\mu}_N = e/(2M_N)$, the nucleon
magneton~\cite{Tanabashi:2018oca,Ramalho:2012pu}. In this work we
take $\mu_{\Sigma^{+}} = 3.112$ and $\mu_{\Sigma^{-}} = -1.479$ with
natural unit~\cite{Tanabashi:2018oca}. In addition, we take
$m_{\rho}$=0.775, $m_{\omega}$=0.782, and $m_{\phi}$=1.019
GeV~\cite{Tanabashi:2018oca}.

In Eqs.~\eqref{eq:f1sigmaps}-\eqref{eq:f2sigmamfv}, $\beta_{\rho}$,
$\beta_{\omega}$, $\beta_{\phi}$, $\alpha_{\phi}$, and
$\alpha_{\rho}$ represent the product of a vector-meson-photon
coupling constant and a $V\Sigma\Sigma$ coupling constant. For the
$V\Sigma\Sigma$ coupling constants, we obtain them through SU(3)
flavor symmetry as in Ref.~\cite{Doring:2010ap},
\begin{eqnarray}
g_{\Sigma \Sigma \omega} &=& g_{B B V} 2 \alpha_{B B V}, \\
g_{\Sigma \Sigma \phi} &=& -g_{B B V} \sqrt{2}\left(2 \alpha_{B B V}-1\right), \\
g_{\Sigma \Sigma \rho} &=& g_{B B V} 2 \alpha_{B B V},
\end{eqnarray}
where $g_{B B V}$ = $g_{N N \rho}$ = 3.20 and $\alpha_{B B V}=1.15$
as used in Refs.~\cite{Machleidt:1987hj,Doring:2010ap}. Then we can
easily get $g_{\Sigma \Sigma \omega}=7.36$, $g_{\Sigma \Sigma
\phi}$=-5.88, $g_{\Sigma \Sigma \rho}=$7.36. While  the tensor
couplings, they are given by
\begin{eqnarray}
f_{\Sigma \Sigma \phi}=f_{N N \rho} \frac{1}{\sqrt{2}}, ~~~f_{\Sigma \Sigma \rho}=f_{N N \rho} \frac{1}{2},
\end{eqnarray}
where $f_{N N \rho}=g_{N N \rho} \kappa_{\rho}$ with
$\kappa_{\rho}$=6.1~\cite{Machleidt:1987hj,Doring:2010ap}. Therefore we can
get $f_{\Sigma \Sigma \phi}$=13.80 and $f_{\Sigma \Sigma
\rho}$=9.76.

In addition, we calculate $V\gamma$ coupling constants following Refs.~\cite{Zhao:2006gw,Huang:2013jda,Huang:2016tcr},
\begin{eqnarray}
\mathcal{L}_{V \gamma}=\sum_{V} \frac{e M_{V}^{2}}{f_{V}} V_{\mu} A^{\mu},\\
\frac{e}{f_{V}}=\left[\frac{3 \Gamma_{V \rightarrow e^{+} e^{-}}}{2 \alpha_{e}\left|\vec{p}_{e}\right|}\right]^{1 / 2}.
\end{eqnarray}
where $\alpha_e = e^{2} /(4 \pi)$=1/137 is the fine-structure
constant and $\vec{p}_e$ is the three momentum of electron in the
rest frame of the vector meson. $\Gamma_{V\to e^+e^-}$ is the
partial decay width of the vector meson decaying into $e^+e^-$ pair.
Then, with the experimental values we get $1/f_{\rho}$=0.200,
$1/f_{\omega}$=0.059, and $1/f_{\phi}$=0.075.

Finally we obtain $\beta_{\rho}$, $\beta_{\omega}$, $\beta_{\phi}$,
$\alpha_{\phi}$ and $\alpha_{\rho}$ through
\begin{eqnarray}
\beta_{V}=g_{\Sigma \Sigma V} \frac{1}{f_{V}}, ~~
\alpha_{V}=f_{\Sigma \Sigma V} \frac{1}{f_{V}},
\end{eqnarray}
which are summarized in Table~\ref{tab:table1}.

\begin{table}[htbp]
    \caption{\label{tab:table1}%
    Parameters used in this work.
    }
\begin{ruledtabular}
    \begin{tabular}{cccc}
        \textrm{Parameter}&
        \textrm{Value}&
        \multicolumn{1}{c}{\textrm{Parameter}}&
        \textrm{Value}\\
        \colrule
        $\beta_{\rho}$ & ~0.736 & $\alpha_{\rho}$ & 0.976\\
        $\beta_{\phi}$ & -0.441 & $\alpha_{\phi}$ & 1.035\\
        $\beta_{\omega}$ &~0.434
    \end{tabular}
\end{ruledtabular}
\end{table}

Next we explain how the large total width of $\rho$ meson
contributions are implemented.~\footnote{We will not consider the
effects from the widths of $\omega$ and $\phi$, since they are so
narrow.} For this purpose, one needs to
replace~\cite{Iachello:1972nu}
\begin{eqnarray}
&& \frac{m_{\rho}^{2}}{m_{\rho}^{2}+Q^{2}} \to \nonumber \\
&& \frac{m_{\rho}^{2}+8 \Gamma_{\rho} m_{\pi} / \pi}{m_{\rho}^{2}+Q^{2}+\left(4 m_{\pi}^{2}+Q^{2}\right) \Gamma_{\rho} \alpha\left(Q^{2}\right) / m_{\pi}},
\label{kuandu}
\end{eqnarray}
where we take $\Gamma_\rho = 149.1$, and a average value of $m_\pi =
138.04$ MeV~\cite{Tanabashi:2018oca}. The function $\alpha(Q^2)$ is
given by
\begin{equation}
\alpha\left(Q^{2}\right)=\frac{2}{\pi}\left(\frac{4 m_{\pi}^{2}+Q^{2}}{Q^{2}}\right)^{1 / 2} \ln \left(\frac{\sqrt{4 m_{\pi}^{2}+Q^{2}}+\sqrt{Q^{2}}}{2 m_{\pi}}\right). \nonumber
\end{equation}

Timelike formfactors can be obtained from the spacelike form factors
by an appropriate analytic continuation. Within the above
ingredients, $g\left(q^{2}\right)$ has the form of an analytical
continuation form,
\begin{equation}
g\left(q^{2}\right) = \left(1-\gamma q^{2}\right)^{-2},
\label{tihuan1}
\end{equation}
where $Q^{2}$ = $-q^2$ = $q^{2} {\rm e}^{{\rm i}\pi}$. We want to
mention that $\gamma$ has positive value, hence $g(q^2)$ will be
divergent at $q^2 = 1/\gamma$. To evade this problem, one can
restrict $\gamma>1 /(4 m_{\Sigma}^{2}) $.

\section{Numerical results and discussion}

In the timelike region, the EMFFs of hyperon $\Sigma^+$ and
$\Sigma^-$ are experimental studied via electron-position
annihilation processes. Under the one-photon exchange approximation,
the total cross section of $e^+e^- \to Y\bar{Y}$, with $Y$ the
$\Sigma^+$ or $\Sigma^-$, can be expressed in terms of the electric
and magnetic form factors $G_{E}$ and $G_{M}$ as~\cite{Denig:2012by}

\begin{equation}
\sigma=\frac{4 \pi \alpha_e^{2} \beta}{3 q^{2}} C_\Sigma \left(\left|G_{M}\left(q^{2}\right)\right|^{2} + \frac{ 2M^2_\Sigma }{q^2}\left|G_{E}\left(q^{2}\right)\right|^{2}\right),
\label{jiemian}
\end{equation}
where $\beta = \sqrt{1-4 M_\Sigma^{2}/q^{2}}$ is a phase-space
factor. $q^2 = s$ is the invariant mass square of the $e^+e^-$
system. In addition, the Coulomb correction factor $C_\Sigma$ is
given by~\cite{Denig:2012by}
\begin{equation}
C(y)=\frac{y}{1-{\rm e}^{-y}},
\end{equation}
with $y = \frac{\alpha \pi}{\beta} \frac{2M_\Sigma}{q}$.

In general, one can easily obtain the effective form factor $G_{\rm
eff}(q^2)$ from the total cross section of $e^+ e^-$ annihilation
process. The effective form factor $G_{\rm eff}(q^2)$ is defined as

\begin{equation}
\left|G_{\mathrm{eff}}\left(q^{2}\right)\right| = \sqrt{\frac{2 \tau\left|G_{M}\left(q^{2}\right)\right|^{2}+\left|G_{E}\left(q^{2}\right)\right|^{2}}{1+2 \tau}}.
\end{equation}

\begin{figure}[htbp]
\centering
\includegraphics[scale=0.5]{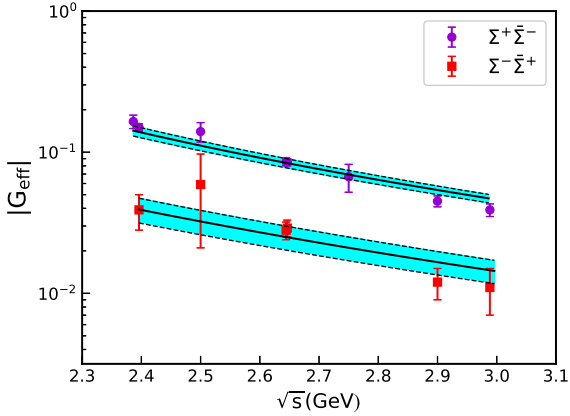}
\caption{The solid curves represent the theoretical results of
$|G_{\rm eff}|$ of the $\Sigma^{+}$ and $\Sigma^{-}$ with the fitted
parameters. The experimental data of $\Sigma^{+}$ and $\Sigma^{-}$
are taken from Ref.~\cite{Ablikim:2020kqp}.} \label{fig:Geff}
\end{figure}

We then perform a $\chi^2$ fit to the experimental data of the effective
form factor $|G_{\rm eff}|$ of $\Sigma^{+}$ and $\Sigma^{-}$ taken
from~\cite{Ablikim:2020kqp}. In the fitting we have only one free
parameter: $\gamma_1$ for $\Sigma^+$, and $\gamma_2$ for $\Sigma^-$.
The fitted parameters are $\gamma_1 = 0.46 \pm 0.01~ {\rm GeV^{-2}}$
and $\gamma_2 = 1.18 \pm 0.13 ~ {\rm GeV^{-2}}$, with $\chi^2/dof =
2.0$ and $1.1$, respectively. The corresponding best-fitting results
for the the effective form factor $|G_{\rm eff}|$ of $\Sigma^{+}$
and $\Sigma^{-}$ in the energy range 2.3864 GeV $< \sqrt{s} <$2.9884
GeV are shown in Fig.~\ref{fig:Geff} with the solid curves. We also
show the theoretical band obtained from the above uncertainties of
the fitted parameters. The numerical results show that we can give a
good description for the experimental data.

\begin{figure}[htbp]
\centering
\includegraphics[scale=0.5]{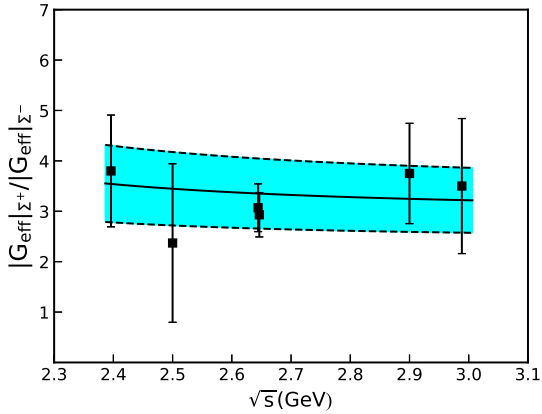}
\caption{Theoretical results for the ratio of $|G_{\rm
eff}|_{\Sigma^{+}}/|G_{\rm eff}|_{\Sigma^{-}}$ compared with the
experimental data taken from Ref.~\cite{Ablikim:2020kqp}.}
\label{fig:ratioGeff}
\end{figure}

From the fitted results of the effective form factors, we can easily
obtain the values of $|G^{\Sigma^+}_{\rm eff}|/|G^{\Sigma^-}_{\rm
eff}|$, which are shown in Fig.~\ref{fig:ratioGeff}. One can see
that the ratio is about three in the energy region of $2.4 <
\sqrt{s} < 3.0$ GeV. The value of 3 is just the ratio of the
incoherent sum of the squared charges of the $\Sigma^+$ and
$\Sigma^-$ valence quarks.

\begin{figure}[htbp]
\centering
\includegraphics[scale=0.5]{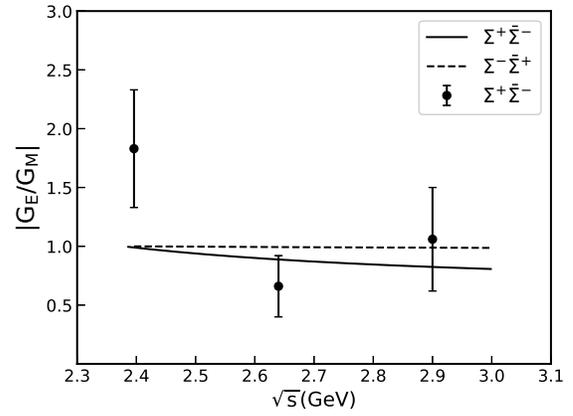}
\caption{The results for the ratio of $|G_{E}/G_{M}|$ of the
$\Sigma^{+}$ and $\Sigma^{-}$. The data are for $\Sigma^{+}$ and
taken from~\cite{Ablikim:2020kqp}. } \label{fig:ratioGEM}
\end{figure}

Through the vector meson dominant model, with the fitted parameter
$\gamma$, one can also easily calculate the ratio of $|G_{E}|$ and
$|G_{M}|$. This ratio is determined to be one at the mass threshold
of a pair of baryon and anti-baryon due to the kinematical
restriction. We shown our theoretical calculations in
Fig.~\ref{fig:ratioGEM}, where the results are obtained with
$\gamma_1 = 0.46$ and $\gamma_2 = 1.18$. It is found that, with the
reaction energy $\sqrt{s}$ increasing, the ratio of $|G_{E}|$ and
$|G_{M}|$ for $\Sigma^{+}$ is slowly decreased, while for the case
of $\Sigma^-$, it is almost flat. Our results here cannot explain
well the experimental data that the ratio is larger than one within
uncertainties close to threshold. This may indicated that there
should be also other contributions around that energy region. For
example, the electromagnetic form factors should be significantly
influenced by the interaction in the final $\Sigma \bar{\Sigma}$
system~\cite{Haidenbauer:2020wyp}. However, since the experimental
and empirical information about the $\Sigma \bar{\Sigma}$ final
state interaction is so limited, we  leave those contributions
to further study when more precise data are available.

Next, we pay attention to the EMFFs in the spacelike region, which
can be straightforwardly calculated with the parameter $\gamma$
determined by the experimental measurements in the timelike region.
Since the parametrization forms shown in
Eqs.~\eqref{eq:f1sigmaps}-\eqref{eq:f2sigmamfv} are valid in the low
$Q^2$ regime, we calculate $G_E$ and $G_M$ below $Q^2 = 3$
$\rm{GeV}^2$, and compare our numerical results with other
calculations.

\begin{figure}[htbp]
\centering
\includegraphics[scale=0.5]{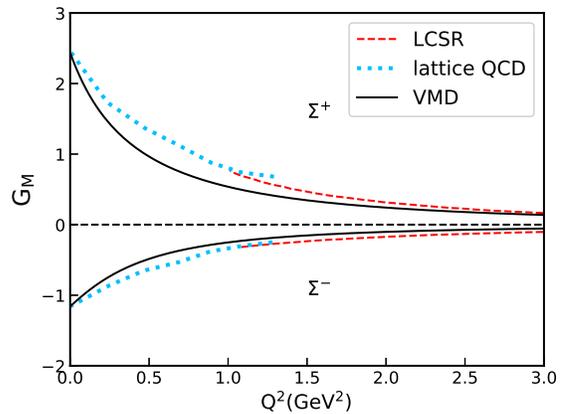}
\caption{The results of the magnetic form factor of $\Sigma^{+}$ and
$\Sigma^{-}$. The blue dotted line are the result of lattice QCD
calculations~\cite{Lin:2008mr}. The red dashed curve are the results
of LCSR calculations~\cite{Liu:2009mb}. }
    \label{fig:GM}
\end{figure}

\begin{figure}[htb]
\centering
\includegraphics[scale=0.5]{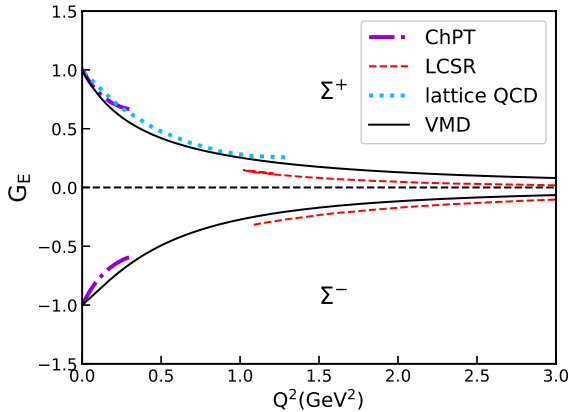}
\caption{Results of the electric form factor $G_{E}$ of the
$\Sigma^{+}$ and $\Sigma^{-}$. The blue dotted line are the result
of lattice QCD~\cite{Lin:2008mr}. The red dashed curve are the
result of LCSR~\cite{Liu:2009mb}. The purple dot dash curve are the
result of chiral perturbation theory~\cite{Kubis:2000aa}. }
\label{fig:GE}
\end{figure}

The numerical results for the $G_M$ and $G_{E}$ obtained with
$\gamma_1 = 0.46$ and $\gamma_2 = 1.18$ are shown in
Fig.~\ref{fig:GM} and \ref{fig:GE}, respectively.~\footnote{To
compare our estimations with other calculations, we convert the unit
of our results into nucleon magneton.} In Fig.~\ref{fig:GM}
predictions from light cone sum rules~\cite{Liu:2009mb} and lattice
QCD calculations~\cite{Lin:2008mr} are also shown for comparing. Our
results for the magnetic form factor of $\Sigma^{+}$ and
$\Sigma^{-}$ are somewhat quantitatively different  from other
theories. Our results for the magnetic form factor of $\Sigma^-$ are
more consistent with other calculations. While for the case of
$\Sigma^-$ electric form factor in Fig.~\ref{fig:GE}, our results
are disagreement with the ChPT and LCSR calculations. However, our
results for the $\Sigma^+$ are closer to the lattice QCD in
Ref.~\cite{Lin:2008mr} and ChPT in the very low $Q^2$ region. It is
expected that these theoretical calculations can be tested by future
experiments on the EMFFs of hyperons $\Sigma^+$ and $\Sigma^-$ and
thus will provide new insight into the complex internal structure of
the baryons.

\section{summary}

In this work, we have investigated the electromagnetic form factors
of the hyperon $\Sigma^{+}$ and $\Sigma^{-}$ within the vector meson
dominance model. The contributions from the $\rho$, $\omega$ and
$\phi$ mesons are taken into account. The model parameters,
$\gamma_1$ and $\gamma_2$, are determined with the BESIII
experimental data on the timelike effective form factors $|G_{\rm
eff}|$ of $\Sigma^{+}$ and ${\Sigma}^{-}$. It is found that the
experimental data can be well reproduced with only one model
parameter. Then, we analytically continue the electromagnetic form
factors to spacelike region and evaluate the spacelike form factors
of $\Sigma^+$ and $\Sigma^-$. The obtained electromagnetic form
factors of the $\Sigma^+$ and $\Sigma^-$ and their ratio are
qualitatively comparable with other model calculations, but slightly
different quantitatively.

Finally, we would like to stress that, the estimations of $\Sigma$
form factors in this work, the $\Lambda$ form factors in
Ref.~\cite{Yang:2019mzq} and proton form factors in
Refs.~\cite{Iachello:1972nu,Iachello:2004aq,Bijker:2004yu} indicate
that the vector meson dominance model is valid to study the
electromagnetic form factors of the baryons, accurate data for the
$e^+ e^- ->$ baryon + anti-baryon reaction can be used to improve
our knowledge of baryon form factors, which are at present poorly
known.

\section*{Acknowledgments}

We thank the anonymous referee for critical comments and suggestions
that are valuable in improving the presentation of the present work.
This work is partly supported by the National Natural Science
Foundation of China under Grants Nos. 12075288, 11735003, and
11961141012.


\begin{thebibliography}{99}

\bibitem{Yang:2019mzq}
Y.~Yang, D.~Y.~Chen and Z.~Lu,
Phys.\ Rev.\ D {\bf 100}, 073007 (2019).

\bibitem{Ramalho:2019koj}
G.~Ramalho, M.~T.~Pe\~{n}a and K.~Tsushima,
Phys.\ Rev.\ D {\bf 101}, 014014 (2020).

\bibitem{Yang:2020rpi}
M.~Yang and P.~Wang,
Phys.\ Rev.\ D {\bf 102}, 056024 (2020).

\bibitem{Ablikim:2020kqp}
M.~Ablikim {\it et al.} [BESIII Collaboration],
arXiv:2009.01404 [hep-ex].

\bibitem{Haidenbauer:2020wyp}
J.~Haidenbauer, U.~G.~Mei{\ss} ner and L.~Y.~Dai,
arXiv:2011.06857 [nucl-th].

\bibitem{Geng:2008mf}
L.~S.~Geng, J.~Martin Camalich, L.~Alvarez-Ruso and M.~J.~Vicente Vacas,
Phys.\ Rev.\ Lett.\  {\bf 101}, 222002 (2008).

\bibitem{Green:2014xba}
J.~R.~Green, J.~W.~Negele, A.~V.~Pochinsky, S.~N.~Syritsyn, M.~Engelhardt and S.~Krieg,
Phys.\ Rev.\ D {\bf 90}, 074507 (2014).
\bibitem{Brodsky:1974vy}
S.~J.~Brodsky and G.~R.~Farrar,
Phys.\ Rev.\ D {\bf 11}, 1309 (1975).
\bibitem{Akhmetshin:2015ifg}
R.~R.~Akhmetshin {\it et al.} [CMD-3 Collaboration],
Phys.\ Lett.\ B {\bf 759}, 634 (2016).
\bibitem{Andreotti:2003bt}
M.~Andreotti {\it et al.},
Phys.\ Lett.\ B {\bf 559}, 20 (2003).
\bibitem{Antonelli:1998fv}
A.~Antonelli {\it et al.},
Nucl.\ Phys.\ B {\bf 517}, 3 (1998).
\bibitem{Ablikim:2019eau}
M.~Ablikim {\it et al.} [BESIII Collaboration],
Phys.\ Rev.\ Lett.\  {\bf 124}, 042001 (2020).
\bibitem{Bardin:1994am}
G.~Bardin {\it et al.},
Nucl.\ Phys.\ B {\bf 411}, 3 (1994).
\bibitem{Bisello:1983at}
D.~Bisello {\it et al.},
Nucl.\ Phys.\ B {\bf 224}, 379 (1983).
\bibitem{Ambrogiani:1999bh}
M.~Ambrogiani {\it et al.} [E835 Collaboration],
Phys.\ Rev.\ D {\bf 60}, 032002 (1999).
\bibitem{Aubert:2005cb}
B.~Aubert {\it et al.} [BaBar Collaboration],
Phys.\ Rev.\ D {\bf 73}, 012005 (2006).
\bibitem{Lees:2013uta}
J.~P.~Lees {\it et al.} [BaBar Collaboration],
Phys.\ Rev.\ D {\bf 88}, 072009 (2013).
\bibitem{Lees:2013ebn}
J.~P.~Lees {\it et al.} [BaBar Collaboration],
Phys.\ Rev.\ D {\bf 87}, 092005 (2013).
\bibitem{Ablikim:2015vga}
M.~Ablikim {\it et al.} [BESIII Collaboration],
Phys.\ Rev.\ D {\bf 91}, 112004 (2015).
\bibitem{Armstrong:1992wq}
T.~A.~Armstrong {\it et al.} [E760 Collaboration],
Phys.\ Rev.\ Lett.\  {\bf 70}, 1212 (1993).
\bibitem{Zhu:2001md}
H.~Zhu {\it et al.} [E93026 Collaboration],
Phys.\ Rev.\ Lett.\  {\bf 87}, 081801 (2001).
\bibitem{Passchier:2001uc}
I.~Passchier {\it et al.},
Phys.\ Rev.\ Lett.\  {\bf 88}, 102302 (2002).
\bibitem{Gayou:2001qd}
O.~Gayou {\it et al.} [Jefferson Lab Hall A Collaboration],
Phys.\ Rev.\ Lett.\  {\bf 88}, 092301 (2002).
\bibitem{Madey:2003av}
R.~Madey {\it et al.} [E93-038 Collaboration],
Phys.\ Rev.\ Lett.\  {\bf 91}, 122002 (2003).
\bibitem{Warren:2003ma}
G.~Warren {\it et al.} [Jefferson Lab E93-026 Collaboration],
Phys.\ Rev.\ Lett.\  {\bf 92}, 042301 (2004).
\bibitem{Becker:1999tw}
J.~Becker {\it et al.},
Eur.\ Phys.\ J.\ A {\bf 6}, 329 (1999).
\bibitem{Bermuth:2003qh}
J.~Bermuth {\it et al.},
Phys.\ Lett.\ B {\bf 564}, 199 (2003).

\bibitem{Golak:2000nt}
J.~Golak, G.~Ziemer, H.~Kamada, H.~Witala and W.~Gloeckle,
Phys.\ Rev.\ C {\bf 63}, 034006 (2001).
\bibitem{Gayou:2001qt}
O.~Gayou {\it et al.},
Phys.\ Rev.\ C {\bf 64}, 038202 (2001).
\bibitem{Ostrick:1999xa}
M.~Ostrick {\it et al.},
Phys.\ Rev.\ Lett.\  {\bf 83}, 276 (1999).
\bibitem{Rohe:1999sh}
D.~Rohe {\it et al.},
Phys.\ Rev.\ Lett.\  {\bf 83}, 4257 (1999).
\bibitem{Chen:2007sa}
D.~Y.~Chen and Y.~B.~Dong,
Commun.\ Theor.\ Phys.\  {\bf 47}, 539 (2007).
\bibitem{Ablikim:2017wlt}
M.~Ablikim {\it et al.} [BESIII Collaboration],
Chin.\ Phys.\ C {\bf 41}, 063001 (2017).
\bibitem{Lin:2008mr}
H.~W.~Lin and K.~Orginos,
Phys.\ Rev.\ D {\bf 79}, 074507 (2009).
\bibitem{Liu:2009mb}
Y.~L.~Liu and M.~Q.~Huang,
Phys.\ Rev.\ D {\bf 79}, 114031 (2009).
\bibitem{Kubis:2000aa}
B.~Kubis and U.~G.~Mei{\ss}ner,
Eur.\ Phys.\ J.\ C {\bf 18}, 747 (2001).
\bibitem{Iachello:1972nu}
F.~Iachello, A.~D.~Jackson and A.~Lande,
Phys.\ Lett.\  {\bf 43B}, 191 (1973).
\bibitem{Iachello:2004aq}
F.~Iachello and Q.~Wan,
Phys.\ Rev.\ C {\bf 69}, 055204 (2004).
\bibitem{Bijker:2004yu}
R.~Bijker and F.~Iachello,
Phys.\ Rev.\ C {\bf 69}, 068201 (2004).
\bibitem{Tanabashi:2018oca}
M.~Tanabashi {\it et al.} [Particle Data Group],
Phys.\ Rev.\ D {\bf 98}, 030001 (2018).

\bibitem{Ramalho:2012pu}
  G.~Ramalho, K.~Tsushima and A.~W.~Thomas,
  J.\ Phys.\ G {\bf 40}, 015102 (2013).
\bibitem{Doring:2010ap}
M.~Doring, C.~Hanhart, F.~Huang, S.~Krewald, U.-G.~Meissner and D.~Ronchen,
Nucl.\ Phys.\ A {\bf 851}, 58 (2011).
\bibitem{Machleidt:1987hj}
R.~Machleidt, K.~Holinde and C.~Elster,
Phys.\ Rept.\  {\bf 149}, 1 (1987).
\bibitem{Zhao:2006gw}
Q.~Zhao, G.~Li and C.~H.~Chang,
Phys.\ Lett.\ B {\bf 645}, 173 (2007).
\bibitem{Huang:2013jda}
Y.~Huang, J.~J.~Xie, X.~R.~Chen, J.~He and H.~F.~Zhang,
Int.\ J.\ Mod.\ Phys.\ E {\bf 23}, 1460002 (2014).
\bibitem{Huang:2016tcr}
Y.~Huang, J.~J.~Xie, J.~He, X.~Chen and H.~F.~Zhang,
Chin.\ Phys.\ C {\bf 40},124104 (2016).
\bibitem{Denig:2012by}
A.~Denig and G.~Salme,
Prog.\ Part.\ Nucl.\ Phys.\  {\bf 68}, 113 (2013).



\end{thebibliography}
 \end{document}